\documentclass[conference]{IEEEtran}
\IEEEoverridecommandlockouts
\usepackage{cite}
\usepackage{amsmath,amssymb,amsfonts}
\usepackage{algorithmic}
\usepackage{graphicx}
\usepackage{textcomp}
\usepackage{xcolor}
\usepackage{subfigure}
\usepackage{tikz}
\usepackage{yhmath}
\usepackage{epstopdf}
\usepackage{graphicx}
\usepackage{hyperref}
\newtheorem{proposition}{\bf Proposition}
\def\BibTeX{{\rm B\kern-.05em{\sc i\kern-.025em b}\kern-.08em
    T\kern-.1667em\lower.7ex\hbox{E}\kern-.125emX}}

\begin{document}

\title{Spatial Bandwidth of Bilateral Near-Field Channels for Linear Large-Scale Antenna Array System}

\author{\IEEEauthorblockN{Zhen Wang\IEEEauthorrefmark{1},
Yijin Pan\IEEEauthorrefmark{1}, Jun-Bo Wang\IEEEauthorrefmark{1}, Yijian Chen\IEEEauthorrefmark{2} and Hongkang Yu\IEEEauthorrefmark{2}}
\IEEEauthorblockA{
\IEEEauthorrefmark{1}National Mobile Communications Research Laboratory,
Southeast University, Nanjing 210096, China.
\\
\IEEEauthorrefmark{2}Wireless Research Institute, ZTE Corporation, Shenzhen 518055, China.
\\
Email: \IEEEauthorrefmark{1}\{wzhen, panyj, jbwang\}@seu.edu.cn,
\IEEEauthorrefmark{2}\{chen.yijian, yu.hongkang\}@zte.com.cn}}

\maketitle

\begin{abstract}
This paper analyzes the spatial bandwidth of line-of-sight (LoS) channels in massive MIMO systems. For the linear large-scale antenna arrays (LSAA) of transceivers placed in random locations in 3D space, a simple but accurate closed-form expression is derived to characterize the spatial bandwidth. Subsequent analysis of the LSAA's spatial bandwidth properties is also provided, leading to the formulation of an approximate expression for the effective degrees of freedom (EDoF) of bilateral near-field channels. Interestingly, as proved in this work, when the transmit and receive arrays are coplanar, with the receive array positioned perpendicular to the axis joining the centroids of the transmit and receive arrays, the EDoF of the LoS channel is found to be approximately maximized.
\end{abstract}


\section{Introduction}
To meet the anticipated demands of future 6G networks in terms of data rates, connection density, latency, reliability, and coverage, array-based communication technologies have exhibited two significant trends: the adoption of Large-Scale Antenna Arrays (LSAA) and the use of higher carrier frequencies, such as the terahertz (THz) band \cite{dardari2020communicating}, \cite{10287956},\cite{akyildiz2018combating}. The key characteristics of LSAA include not only an increased number of antennas but also an expanded array aperture, enabling LSAA communication systems to capture richer spatial domain information \cite{lu2021communicating}. In the THz band, the high carrier frequency offers abundant bandwidth resources; however, due to higher penetration losses, multipath propagation is considerably weakened, which makes Line-of-Sight (LoS) array communication increasingly important \cite{tang2023line}.

The conventional view holds that LoS environments are unfavorable for multiple-input multiple-output (MIMO) communication, as MIMO channels in such scenarios cannot provide spatial multiplexing gains \cite{tse2005fundamentals}. However, the use of LSAA positions users predominantly in the near-field radiation of access points, challenging the traditional assumption of plane-wave propagation. Instead, a more general non-uniform spherical wave propagation model is required to accurately characterize phase and amplitude variations within the array \cite{lu2024tutorial}, \cite{tang2024joint}, \cite{pan2023ris}. Within the near-field region, spherical waves exhibit nonlinear changes in phase and power levels across each link, thereby increasing the rank of the channel matrix \cite{liu2024near}. This significantly enhances spatial degrees of freedom (DoF) in MIMO channels in near-field LoS environments, even in LoS conditions.

Significant progress has been made in understanding the spatial DoF, particularly the Effective Degrees of Freedom (EDoF) \cite{sun2024degree}, \cite{yuan2021electromagnetic}, \cite{decarli2021communication}. For example, the authors in \cite{sun2024degree} proposed a DOF analysis method for Holographic MIMO systems based on a mutual-coupling-compliant channel model. 
The authors in \cite{yuan2021electromagnetic} conducted a numerical analysis of the EDoF for near-field spatially-discrete MIMO channels, demonstrating that the number of EDoF is inversely proportional to the propagation distance. In \cite{decarli2021communication}, the authors derived an approximate expression for the number of EDoF by leveraging diffraction theory and the energy focusing properties of near-field beamforming, indicating that MIMO-like communication can be achieved even under LoS conditions without relying on multipath propagation. In fact, for array communication in LoS environments, the channel is completely determined by the geometric characteristics of the transmitter and receiver arrays, including their size, position, and orientation. However, these studies offer limited insights into the spatial DoF from the perspective of array geometry.

Recently, \cite{franceschetti2017wave} proposed a general formula for calculating achievable spatial DoF using spatial bandwidth. \cite{ding2022degrees} obtained closed-form approximations for each orthogonal reception direction by performing an orthogonal decomposition of the receiving directions and analyzed the effects of array position and rotation on achievable spatial DoF. In \cite{ding2023spatial}, a dual-slope asymptotic expression was used to approximate the spatial bandwidth at the center of the receiving array, and this expression was utilized to analyze the DoF performance of the channel. However, both \cite{ding2022degrees} and \cite{ding2023spatial}  yield only approximate expressions for spatial bandwidth, with derivations that are relatively complex and lacking more intuitive geometric insights. Additionally, several interesting properties of spatial bandwidth remain unexplored.

To address these issues, this paper derives a simple but accurate closed-form expression for the spatial bandwidth of a linear LSAA. This expression highlights the effects of the transmit and receive array apertures, their relative positioning, and orientation on spatial bandwidth and EDoF. Notably, the EDoF of the LoS channel reaches its maximum when the transmit and receive arrays are coplanar with the receive array oriented perpendicularly to the line joining their centroids, and it attains its minimum when the arrays are perpendicular to each other.

% To address these issues, in this paper, we derive an simple but accuarate closed-form expression for the spatial bandwidth for a linear LSAA. 
% This expression elucidates the impact of the apertures of the transmitting and receiving arrays, as well as their relative positions and orientations, on spatial bandwidth, revealing several interesting properties. 
% % Furthermore, using the derived spatial bandwidth expression, we propose two methods for calculating the achievable DoF of the channel and reveal the influencing factors of the EDoF of the LoS channel from a geometric perspective.

\section{Spatial Bandwidth Analysis for 3D linear LSAA Communications}
\subsection{Closed-Form Expression for Local Spatial Bandwidth}
\begin{figure}[h]
\centering
\includegraphics[width=6cm]{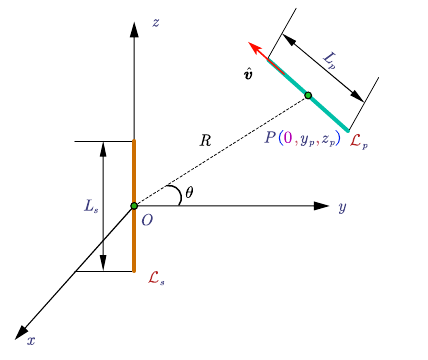}
\caption{System model.}
\label{system_model}
\end{figure}

As illustrated in Fig.\ref{system_model}, we consider both the transmitter and receiver as linear LSAAs, denoted by $\mathcal{L}_p$ and $\mathcal{L}_s$, respectively, with dimensions $L_p$ and $L_q$. We examine the LoS wireless channel between them under 3D free-space propagation conditions, assuming that $\mathcal{L}_p$ has perfect and interference-free perception of the electric field \cite{ding2022degrees}.

The transmitting array $\mathcal{L}_s$ is centered at the origin $O$ and oriented along the $z$-axis. The point $P$ represents the center of the receiving array $\mathcal{L}_p$, with coordinates $\boldsymbol{p}_0 = (x_p, y_p, z_p)$. The orientation of the $\mathcal{L}_p$ is given by the unit vector $\hat{\boldsymbol{v}} = \left[ v_x, v_y, v_z \right]^T$, where $v_z \ge 0$ and $\left \| \hat{\boldsymbol{v}} \right \| = 1$. Under the assumption of an ideal isotropic point source, rotating $\mathcal{L}_p$ around the $z$-axis does not affect the electric field observed by it.

For simplicity, we derive the expression for the local spatial bandwidth at the center of $\mathcal{L}_p$. It should be noted that the local spatial bandwidth at any point in 3D space can be obtained using the same method.

\begin{figure}[h]
\centering
\includegraphics[width=6cm]{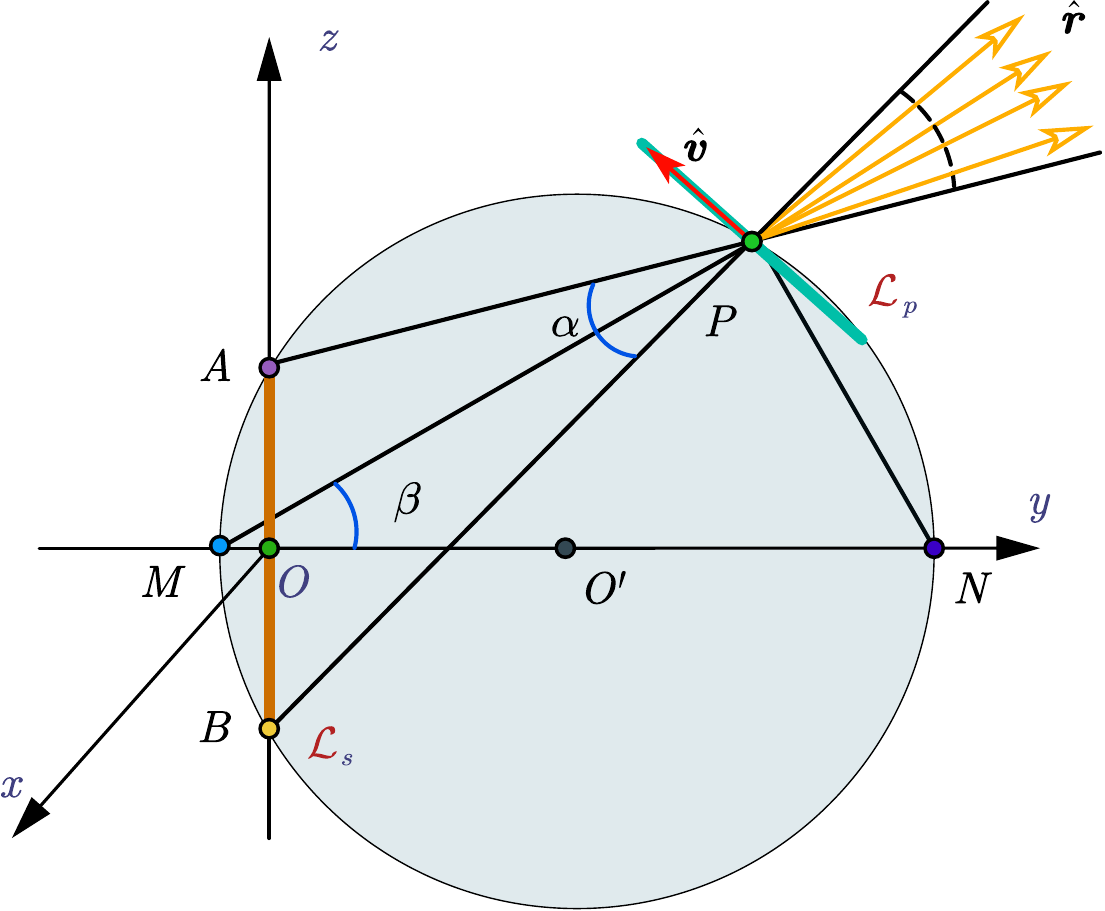}
\caption{The range of variation of the propagation direction vector $\hat{\boldsymbol{r}}$.}
\label{local_BW_figure}
\end{figure}

Without loss of generality, suppose that $P$ lies in the first quadrant of the $yOz$ plane (including the boundary), i.e., $x_p = 0$, $y_p \ge 0$, and $z_p \ge 0$. Furthermore, $\mathcal{L}_s$ and $\mathcal{L}_p$ do not intersect. Let $R = \left| OP \right|$ denote the distance between the origin and $P$, and let $\theta$ be the angle between the vector $\boldsymbol{p}_0$ and the positive $y$-axis $\hat{\boldsymbol{y}} = \left[ 0, 1, 0 \right]^T$. Therefore, the position of $\mathcal{L}_p$ can be described as $\boldsymbol{p}_0 = \left( 0, R\cos \theta, R\sin \theta \right)$. Unless otherwise specified, all subsequent derivations and simulations are based on this configuration.

According to electromagnetic (EM) propagation theory, we define:

\textbf{Definition 1.} The \textit{spatial frequency} of the wave component generated by the source $\boldsymbol{s} \in \mathcal{L}_s$, measured at $\boldsymbol{p}_0 \in \mathcal{L}_p$, as it moves in the direction of $\hat{\boldsymbol{v}}$, is given by
\begin{equation}
\label{spaitial_frequency_def}
f_{\hat{\boldsymbol{v}}}\left( \boldsymbol{p}_0,\boldsymbol{s} \right) =k_0\hat{\boldsymbol{r}}^T\left( \boldsymbol{p}_0,\boldsymbol{s} \right) \hat{\boldsymbol{v}},
\end{equation}
where $k_0 = 2\pi/\lambda$ represents the wavenumber, and $\lambda$ is the wavelength of the EM wave. The vector $\hat{\boldsymbol{r}}(\boldsymbol{p}_0, \boldsymbol{s}) = (\boldsymbol{p}_0 - \boldsymbol{s}) / \| \boldsymbol{p}_0 - \boldsymbol{s} \|$ denotes the propagation direction, as depicted in Fig. \ref{local_BW_figure}.
From (\ref{spaitial_frequency_def}), it is evident that the spatial frequency $f_{\hat{\boldsymbol{v}}}\left( \boldsymbol{p}_0, \boldsymbol{s} \right)$ can be expressed as the inner product of two unit vectors. Consequently, $ -k_0\le f_{\hat{\boldsymbol{v}}}\left( \boldsymbol{p}_0, \boldsymbol{s} \right) \le k_0$. It is also clear that different points $\boldsymbol{s}' \in \mathcal{L}_s$ will generate distinct wave components at $\boldsymbol{p}_0$.
Accordingly, the \textit{local spatial bandwidth} is defined as follows:

\textbf{Definition 2.} The \textit{local spatial bandwidth} at point $\boldsymbol{p}_0$ is defined as the difference between the maximum and minimum spatial frequencies of all wave components radiated by $\mathcal{L}_s$, given by
\begin{equation}
\label{local_BW_def}
\omega _{\hat{\boldsymbol{v}}}\left( \boldsymbol{p}_0,\mathcal{L} _s \right) =k_0\left( \max_{\boldsymbol{s}\in \mathcal{L} _s} \hat{\boldsymbol{r}}^T\left( \boldsymbol{p}_0,\boldsymbol{s} \right) \hat{\boldsymbol{v}}-\min_{\boldsymbol{s}\in \mathcal{L} _s} \hat{\boldsymbol{r}}^T\left( \boldsymbol{p}_0,\boldsymbol{s} \right) \hat{\boldsymbol{v}} \right).
\end{equation}
From (\ref{local_BW_def}), we observe that $0 \le \omega_{\hat{\boldsymbol{v}}}\left( \boldsymbol{p}_0, \mathcal{L}_s \right) \le 2k_0$. It should be noted that $\omega_{\hat{\boldsymbol{v}}}\left( \boldsymbol{p}_0, \mathcal{L}_s \right)$ is influenced by the position of the observation point $\boldsymbol{p}_0$ and the orientation $\hat{\boldsymbol{v}}$ of $\mathcal{L}_p$, but remains independent of the dimension $L_p$.

To derive a closed-form expression for $\omega_{\hat{\boldsymbol{v}}}\left( \boldsymbol{p}_0, \mathcal{L}_s \right)$, the key idea is to characterize the region of variation of the propagation direction vector $\hat{\boldsymbol{r}}$ at different positions $\boldsymbol{p}_0$.
As shown in Fig. \ref{local_BW_figure}, denote the endpoints of $\mathcal{L}_s$ as $A$ and $B$, and connect $AP$ and $BP$. Let $\alpha = \angle APB$ represent the angle subtended by $P$ with respect to $A$ and $B$, satisfying $0 < \alpha < \pi$.
Construct the circumcircle of triangle $APB$, intersecting the $y$-axis at points $M$ and $N$, with $O'$ as its center. Since the $y$-axis acts as the perpendicular bisector of segment $AB$, $O'$ lies on this axis. Connect $MP$, noting that $MP$ serves as the bisector of angle $\alpha$. Define $\hat{\boldsymbol{v}}_\mathrm{{MP}} = \overrightarrow{MP} / \left| MP \right|$. Similarly, connect $NP$ and define $\hat{\boldsymbol{v}}_\mathrm{{NP}} = \overrightarrow{NP} / \left| NP \right|$.
Given that $MN$ represents the diameter of circle $O'$, it follows that $MP \perp NP$. Let $\beta = \angle PMN$, which can be determined using simple geometric relationships, with $0 \le \beta \le \pi/2$.

The propagation direction vector $\hat{\boldsymbol{r}}$ can be described in spherical coordinates as follows:
\begin{equation}
    \label{r_sph_coor}
    \hat{\boldsymbol{r}}\left( \boldsymbol{p}_0,\boldsymbol{s} \right) \triangleq \hat{\boldsymbol{r}}\left( \gamma \right) =\left[ 0,\cos \gamma ,\sin \gamma \right] ^T,
\end{equation}
where $\gamma$ represents the angle between $\hat{\boldsymbol{r}}$ and $\hat{\boldsymbol{y}}$, satisfying $-\alpha /2 + \beta \le \gamma \le \alpha /2 + \beta$. Since $A$, $P$, and $B$ are all situated in the $yOz$ plane, the $x$-axis component of $\hat{\boldsymbol{r}}$ is zero.
We also note that $(\alpha, \beta)$ determines the position of $\boldsymbol{p}_0$, meaning that the range of variation of $\hat{\boldsymbol{r}}$ is related to $\boldsymbol{p}_0$.
We use two angles, $(\alpha, \beta)$, to simultaneously describe the position of $\boldsymbol{p}_0$ and the range of variation of the propagation direction vector $\hat{\boldsymbol{r}}$.

Similarly, we express the orientation $\hat{\boldsymbol{v}}$ of $\mathcal{L}_p$ using spherical coordinates: 
\begin{equation}
    \label{v_sph_coor}
    \hat{\boldsymbol{v}}\left( \psi ,\varphi \right) =\left[ \cos \psi ,\sin \psi \cos \varphi ,\sin \psi \sin \varphi \right] ^T,
\end{equation}
where $\psi$ denotes the polar angle of $\hat{\boldsymbol{v}}$ with respect to the positive $x$-axis $\hat{\boldsymbol{x}} = \left[ 1, 0, 0 \right]^T$, satisfying $0 \le \psi \le \pi$. The angle $\varphi$ represents the azimuth angle of the projection $\hat{\boldsymbol{v}}_{\mathrm{proj}}$ of $\hat{\boldsymbol{v}}$ onto the $yOz$ plane, relative to $\hat{\boldsymbol{y}}$, with $0 \le \varphi \le \pi$.

Based on (\ref{r_sph_coor}) and (\ref{v_sph_coor}), the spatial frequency $f_{\hat{\boldsymbol{v}}}\left( \boldsymbol{p}_0, \boldsymbol{s} \right)$ can be reformulated using angles $\left( \psi, \varphi, \gamma \right)$ as:
\begin{equation}
\label{re_spatial_frequency}
\begin{aligned}
f_{\hat{\boldsymbol{v}}}\left(\boldsymbol{p}_0, \boldsymbol{s}\right) & =\hat{\boldsymbol{r}}^T(\gamma) \hat{\boldsymbol{v}}(\psi, \varphi) \\
& =k_0[\sin \psi \cos \gamma \cos \varphi+\sin \psi \sin \gamma \sin \varphi] \\
& =k_0 \sin \psi \cos (\gamma-\varphi) \triangleq f(\psi, \varphi, \gamma).
\end{aligned}
\end{equation}
Let the maximum and minimum values of the spatial frequency received at point $P$, when $\mathcal{L}_p$ is oriented as $\hat{\boldsymbol{v}}\left( \psi, \varphi \right)$, be denoted as $F^{\max}\left( \psi, \varphi ; \alpha, \beta \right)$ and $F^{\min}\left( \psi, \varphi ; \alpha, \beta \right)$, respectively, as follows:
\begin{equation}
    \label{F_max_def}
F^{\max}\left( \psi ,\varphi ;\alpha ,\beta \right) \triangleq \underset{-\alpha /2+\beta \le \gamma \le \alpha /2+\beta}{\max}f\left( \psi ,\varphi ,\gamma \right), 
\end{equation}
\begin{equation}
    \label{F_min_def}
F^{\min}\left( \psi ,\varphi ;\alpha ,\beta \right) \triangleq \underset{-\alpha /2+\beta \le \gamma \le \alpha /2+\beta}{\min}f\left( \psi ,\varphi ,\gamma \right). 
\end{equation}
According to (\ref{local_BW_def}), the local spatial bandwidth received at $\boldsymbol{p}_0$ can be expressed as
\begin{equation}
    \label{w_def_F}
\omega \left( \psi ,\varphi ;\alpha ,\beta \right) =F^{\max}\left( \psi ,\varphi ;\alpha ,\beta \right) -F^{\min}\left( \psi ,\varphi ;\alpha ,\beta \right).
\end{equation}

% \textbf{Proposition 1.} For all $\alpha, \beta$, $\omega$ is a periodic function of $\varphi$, given by
% % \begin{lemma}
% % Given two line segments whose lengths are \(a\) and \(b\) respectively there 
% % is a real number \(r\) such that \(b=ra\).
% % \end{lemma}

It can be easily verified that for any $(\alpha, \beta)$, $\omega \left( \psi ,\varphi ;\alpha ,\beta \right)$ is a periodic function of $\varphi$ with a period of $\pi$, given by:
\begin{equation}
\label{periodic_function}
\omega \left( \psi ,\varphi ;\alpha ,\beta \right) =\omega \left( \psi ,\varphi +\pi ;\alpha ,\beta \right). 
\end{equation}
This implies that when analyzing the function $\omega \left( \psi, \varphi ; \alpha, \beta \right)$ over $\psi \in \left[ 0, \pi \right]$ and $\varphi \in \left[ 0, \pi \right]$, it suffices to consider its values over $\psi \in \left[ 0, \pi \right]$ and $\varphi \in \left[ \mu, \mu + \pi \right]$, where $\mu$ can range over any real number in $\left[ 0, \pi \right]$.
This property can simplify the analysis of $\omega \left( \psi, \varphi ; \alpha, \beta \right)$.
By setting $\mu = \beta$, we have $\varphi \in \left[ \beta, \beta + \pi \right]$. Define $\varphi' \triangleq \varphi - \beta$, which represents the angle of the projection $\hat{\boldsymbol{v}}_{\mathrm{proj}}$ of $\hat{\boldsymbol{v}}$ onto the $yOz$ plane relative to $\hat{\boldsymbol{v}}_\mathrm{{MP}}$.
Next, we derive the closed-form expression of $\omega \left( \psi, \varphi' ; \alpha, \beta \right)$ over $\psi \in \left[ 0, \pi \right]$ and $\varphi' \in \left[ 0, \pi \right]$.

Given $\sin \psi \ge 0$ and equations (\ref{F_max_def}) and (\ref{F_min_def}), we have
\begin{equation}
\label{F_max_cf_expr}
\begin{aligned}
& F^{\max }\left(\psi, \varphi^{\prime} ; \alpha, \beta\right) \\
& \quad=\left\{\begin{array}{l}
k_0 \sin \psi, \varphi^{\prime} \in[0, \alpha / 2], \\
k_0 \sin \psi \cos \left(\alpha / 2-\varphi^{\prime}\right), \varphi^{\prime} \in(\alpha / 2, \pi],
\end{array}\right.
\end{aligned}
\end{equation}
and
\begin{equation}
\label{F_min_cf_expr}
\begin{aligned}
& F^{\min }\left(\psi, \varphi^{\prime} ; \alpha, \beta\right) \\
& \quad=\left\{\begin{array}{l}
k_0 \sin \psi \cos \left(-\alpha / 2-\varphi^{\prime}\right), \varphi^{\prime} \in[0, \pi-\alpha / 2), \\
-k_0 \sin \psi, \varphi^{\prime} \in[\pi-\alpha / 2, \pi].
\end{array}\right.
\end{aligned}
\end{equation}
From equations (\ref{w_def_F}), (\ref{F_max_cf_expr}), and (\ref{F_min_cf_expr}), we obtain the following expression for the local spatial bandwidth at $\boldsymbol{p}_0$:
\begin{equation}
\label{w_closed_form_expr}
\begin{aligned}
& \omega\left(\psi, \varphi^{\prime} ; \alpha, \beta\right) \\
& =\left\{\begin{array}{l}
k_0 \sin \psi\left[1-\cos \left(-\alpha / 2-\varphi^{\prime}\right)\right], 0 \leq \varphi^{\prime} \leq \alpha / 2, \\
2 k_0 \sin \psi \sin (\alpha / 2) \sin \varphi^{\prime}, \alpha / 2<\varphi^{\prime}<\pi-\alpha / 2, \\
k_0 \sin \psi\left[1+\cos \left(\alpha / 2-\varphi^{\prime}\right)\right], \pi-\alpha / 2 \leq \varphi^{\prime} \leq \pi.
\end{array}\right.
\end{aligned}
\end{equation}

\textbf{Remark 1:} It is noteworthy that, different from the approximate expressions derived in previous works \cite{ding2022degrees} and \cite{ding2023spatial}, (\ref{w_closed_form_expr}) accurately computes the local spatial bandwidth at any point in the 3D space. Moreover, (\ref{w_closed_form_expr}) provides an intuitive geometric interpretation of the factors influencing local spatial bandwidth. Specifically, the expression is characterized by two sets of angle parameters: the first set $\left( \psi, \varphi' \right)$ relates to the orientation $\hat{\boldsymbol{v}}$ of $\mathcal{L}_s$, and the second set $\left( \alpha, \beta \right)$ relates to the position $\boldsymbol{p}_0$ of $\mathcal{L}_p$. Additionally, it is observed that in (\ref{w_closed_form_expr}), $\psi$ and $\varphi'$ are decoupled, allowing us to independently study the effects of the elevation angle and the azimuth angle on the received spatial bandwidth. This expression enables the fast and accurate determination of spatial bandwidth at various points in space, thereby accelerating system-level simulations.

% \begin{figure*}[t]
% \centering
% \subfloat[Case I]{\includegraphics[width=2in]{P_0_100_theta_0.pdf}%
% \label{fig_first_case}}
% \hfil
% \subfloat[Case II]{\includegraphics[width=2in]{P_0_50_theta_0.pdf}%
% \label{fig_second_case}}
% \subfloat[Case III]{\includegraphics[width=2in]{P_0_100_0_theta_45.pdf}%
% \label{fig_third_case}}
% \caption{Simulation results for the network.}
% \label{fig_sim}
% \end{figure*}

\begin{figure}[!htbp]
\centering
\subfigure[$(R,\theta)=(100 \lambda, 0)$]{
	\label{localbw_sim_A} %% label for first subfigure
	\includegraphics[width=5.5cm]{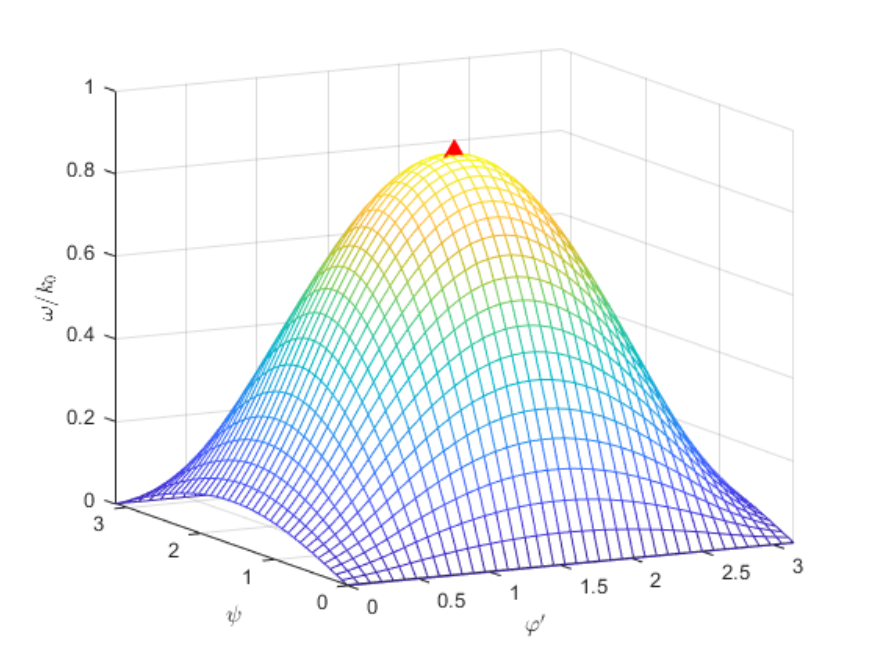}}
\hspace{0.5in}
\subfigure[$(R,\theta)=(100 \lambda, \pi/4)$]{
	\label{localbw_sim_B} %% label for second subfigure
	\includegraphics[width=5.5cm]{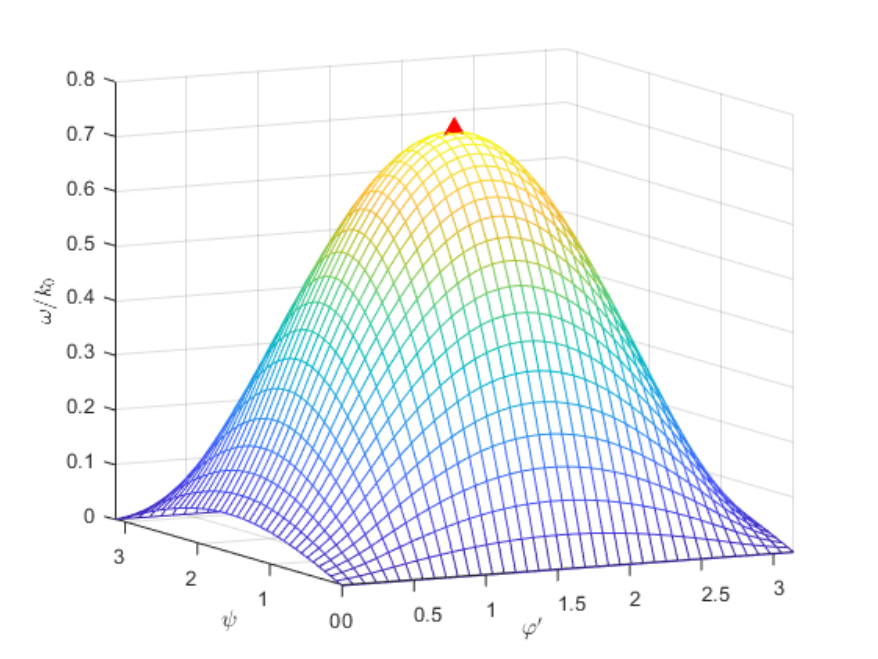}}
\caption{The relationship between local spatial bandwidth and receiving orientation when the observation point is located at different positions.}
\label{figure_local_BW}
\end{figure}

As shown in Fig. \ref{figure_local_BW}, we plot the local spatial bandwidth $\omega$ as a function of the orientation $(\psi, \varphi')$ of the receiving array, based on (\ref{w_closed_form_expr}). The length of the transmitting array is $L_s = 100\lambda$, and the position of $P$ is described by $\left( R, \theta \right)$.  Note that here, $\varphi'$ is the angle between the projection $\hat{\boldsymbol{v}}_{\mathrm{proj}}$ of $\hat{\boldsymbol{v}}$ onto the $yOz$ plane and $\hat{\boldsymbol{v}}_{\mathrm{MP}}$. 

From Fig. \ref{localbw_sim_A}, it is observed that the local spatial bandwidth $\omega$ at $P$ initially increases and then decreases as $\psi$ varies from $0$ to $\pi$. For all $\varphi' \in \left[ 0, \pi \right]$, $\omega$ reaches its maximum value when $\psi = \pi / 2$. Notably, at this point, the component of $\hat{\boldsymbol{v}}$ along $\hat{\boldsymbol{x}}$ is zero, indicating that $\mathcal{L}_p$ and $\mathcal{L}_s$ are coplanar. When $\psi$ is either $0$ or $\pi$, i.e., when $\mathcal{L}_p$ is perpendicular to $\mathcal{L}_s$, $\omega$ attains its minimum value of zero.
Similarly, as $\varphi'$ changes from $0$ to $\pi$, the local spatial bandwidth $\omega$ at $P$ also increases initially and then decreases. For all $\psi \in \left[ 0, \pi \right]$, $\omega$ achieves its maximum value when $\varphi' = \pi / 2$.
The maximum value of $\omega$ is attained when $\left( \psi, \varphi' \right) = \left( \pi / 2, \pi / 2 \right)$.
In this configuration, the optimal receive array orientation that maximizes the local spatial bandwidth of $P$, denoted as $\hat{\boldsymbol{v}}^{\mathrm{opt}}$, satisfies $\hat{\boldsymbol{v}}^{\mathrm{opt}} = \hat{\boldsymbol{v}}_{NP}$.

From Fig. \ref{localbw_sim_B}, it is evident that the maximum value of the local spatial bandwidth varies with different observation azimuth angles $\theta$, while the trend of $\omega$ remains consistent with that observed in Fig. \ref{localbw_sim_A}. This consistency arises from our use of a relative coordinate method to describe the orientation of $\mathcal{L}_p$, which facilitates a better understanding of the spatial bandwidth properties.

% \begin{figure}[h]
%     \centering
%     \includegraphics[width=6cm]{P_0_100_theta_0.pdf}
%     \caption{Caption}
%     \label{localbw_sim}
% \end{figure}

\subsection{Maximum of the Local Spatial Bandwidth}
From (\ref{w_closed_form_expr}), we can further explore the relationship between the local spatial bandwidth and the position parameters $(\alpha, \beta)$ of $\mathcal{L}_p$. Specifically, we investigate the maximum value of the local spatial bandwidth at a given location, defined as
\begin{equation}
\omega ^{\max}\left( \alpha ,\beta \right) =\max_{\psi ,\varphi ^{\prime}} \,\,\omega \left( \psi ,\varphi ^{\prime};\alpha ,\beta \right). 
\end{equation}

\begin{proposition}
\label{prop2}
For any point $P$ in 3D space, the local spatial bandwidth $\omega$ at $P$ reaches its maximum value if and only if the receiving direction $\hat{\boldsymbol{v}}$ satisfies $\hat{\boldsymbol{v}} = \pm \hat{\boldsymbol{v}}_{NP}$. The maximum value $\omega^{\max}(\alpha,\beta)$ is given by
\begin{equation}
\omega^{\max}(\alpha,\beta) = 2k_0 \sin \left( \alpha /2 \right).
\end{equation}
\end{proposition}
\begin{IEEEproof}[Proof]
	see Appendix A.
\end{IEEEproof}

\textbf{Remark 2:} According to Proposition 1, the maximum local spatial bandwidth $\omega^{\max}$ at $P$ depends solely on the angle $\alpha$ subtended by $P$ and the two ends of $\mathcal{L}_s$. The larger the angle $\alpha$, the greater the $\omega^{\max}$. The orientation $\hat{\boldsymbol{v}}$ that corresponds to this maximum value $\omega^{\max}$, however, depends on the angle $\beta$. Furthermore, based on geometric relationships, any point $P'$ on the circular arc $\wideparen{APB}$ has the same maximum local spatial bandwidth $\omega^{\max}$.

\begin{figure}
    \centering
    \includegraphics[width=7.5cm]{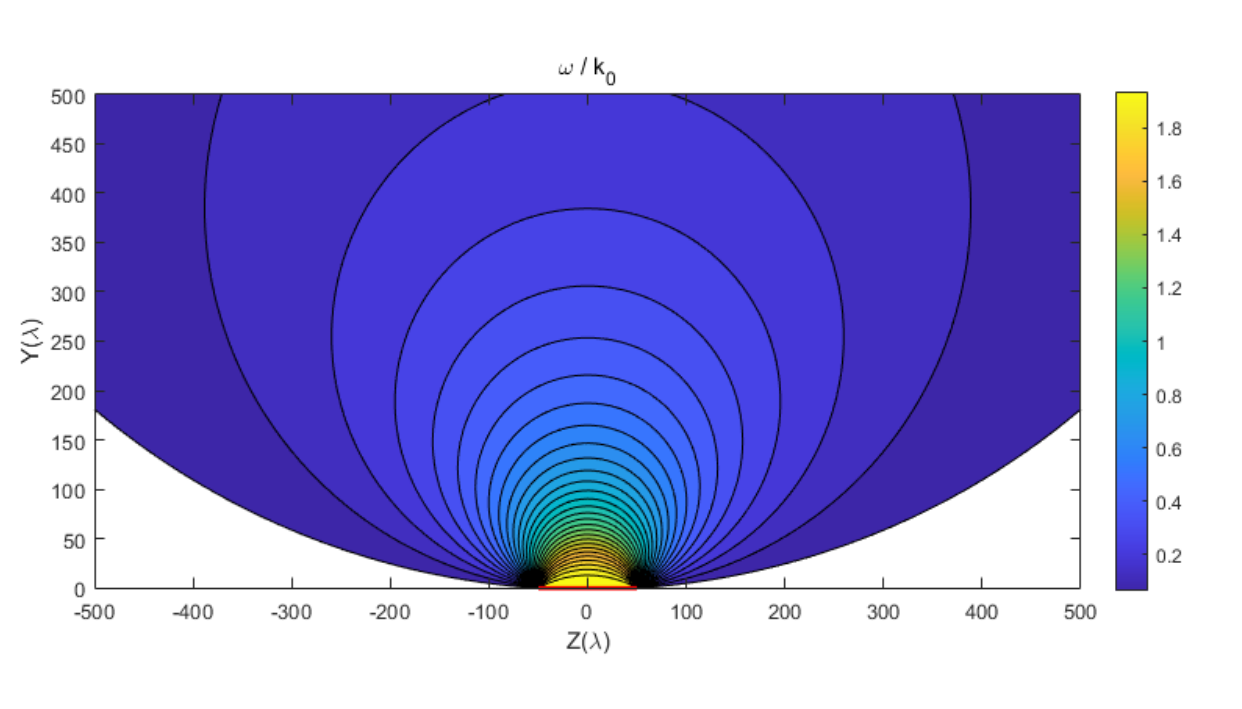}
    \caption{The maximum of local spatial bandwidth at different observation points.}
    \label{max_BW_figure}
\end{figure}

Figure \ref{max_BW_figure} illustrates the ratio of the maximum local spatial bandwidth to the wavenumber $k_0$ for various points in the $yOz$-plane when $L_s = 100 \lambda$. The black lines depict the contour lines, and the red line represents the position of $\mathcal{L}_s$. For a fixed observation angle $\theta$, $\omega^{\max}$ decreases as $R$ increases, and the rate of decrease of $\omega^{\max}$ slows down with increasing $R$. When $R$ is fixed and $R > L_s/2$, $\omega^{\max}$ increases as $\theta$ approaches $0$. Additionally, it is observed that the contour lines of $\omega^{\max}$ form circles with $\mathcal{L}_s$ as their chord. This observation is consistent with the analysis in Remark 2.

% \begin{figure}
% \centering
% \label{syestem_model}
% \includegraphics[width=6cm]{system_model.pdf}
% \caption{System model.}
% \end{figure}

% \section*{Acknowledgment}

% The preferred spelling of the word ``acknowledgment'' in America is without 
% an ``e'' after the ``g''. Avoid the stilted expression ``one of us (R. B. 
% G.) thanks $\ldots$''. Instead, try ``R. B. G. thanks$\ldots$''. Put sponsor 
% acknowledgments in the unnumbered footnote on the first page.

\section{K Number and Effective Degrees of Freedom}
\subsection{K Number}
With (\ref{w_closed_form_expr}), the number of samples provided by non-redundant Nyquist sampling (i.e., denoted as K number \cite{franceschetti2017wave}), can be computed using numerical or approximate methods. Many studies suggest that $K$ number can approximate the EDoF for a MIMO channel \cite{ding2022degrees}, \cite{ding2023spatial}. According to \cite{ding2022degrees}, for a 1D receiving array, the $K$ number can be calculated as
\begin{equation}
\label{K_def_expr}
K=\frac{1}{2\pi}\int_{-L_p/2}^{L_p/2}{\omega _{\hat{\boldsymbol{v}}}\left( l_p,\mathcal{L} _s \right) \mathrm{d}l_p},
\end{equation}
where $\omega _{\hat{\boldsymbol{v}}}\left( l_p,\mathcal{L}_s \right)$ denotes the local spatial bandwidth at the receiving array $\mathcal{L}_p$ located at $\boldsymbol{p}_0 + l_p \hat{\boldsymbol{v}}$ when $\mathcal{L}_p$ is oriented along $\hat{\boldsymbol{v}}$. Since we have obtained a closed-form expression for the local spatial bandwidth at any point in 3D space, the K number can be approximated using numerical method. 

When $L_p$ is small relative to $R$, the local spatial bandwidth at different points across the $\mathcal{L}_p$ is approximately equal. Therefore, the local spatial bandwidth at the center of $\mathcal{L}_p$, denoted as $\omega _{\hat{\boldsymbol{v}}}\left( \boldsymbol{p}_0, \mathcal{L}_s \right)$, can serve as an approximation for the local spatial bandwidth at all points on $\mathcal{L}_p$. Therefore, the K number can be approximated as:
\begin{equation}
\label{K2}
K_{2}^{}=\frac{L_p}{2\pi}\omega _{\hat{\boldsymbol{v}}}\left( \boldsymbol{p}_0,\mathcal{L} _s \right). 
\end{equation}
Under this condition, the $K$ number is proportional to the local spatial bandwidth $\omega _{\hat{\boldsymbol{v}}}\left( \boldsymbol{p}_0, \mathcal{L}_s \right)$ at the center of the array and the length $L_p$ of the receiving array. Thus, the orientation $\hat{\boldsymbol{v}}$ that maximizes the $K$ number of the receiving array $\mathcal{L}_p$ is approximately equal to the orientation $\hat{\boldsymbol{v}}$ that maximizes the local spatial bandwidth at the center of $\mathcal{L}_p$. According to Proposition 1, the maximum value of the $K$ number can be approximated as
\begin{equation}
\label{K2_max}
K_{2}^{\max} = \frac{k_0L_p}{\pi}\sin \left( \alpha /2 \right). 
\end{equation}
The accuracy of this approximation will be validated through simulation results.

\begin{figure}[h]
    \centering
    \includegraphics[width=5.5cm]{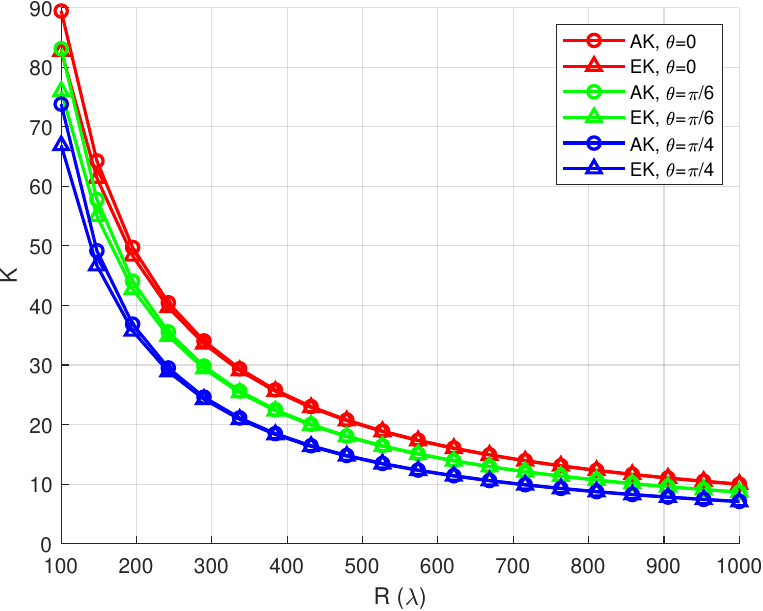}
    \caption{The maximum K number corresponding to the receiving array at different positions.}
    \label{max_K_figure}
\end{figure}

Figure \ref{max_K_figure} illustrates the relationship between the position of the receiving array $\mathcal{L}_p$ and its maximum $K$ number when $L_s = 100 \lambda$, $L_p = 100 \lambda$, and $ \lambda = 0.01$ m, where the center position of $\mathcal{L}_p$ is described by $(R, \theta)$. ``AK" represents the maximum value of the K number calculated using (\ref{K2_max}). ``EK" denotes the K number calculated using numerical methods and then maximized by traversing all possible receiving directions $\hat{\boldsymbol{v}}$.
As shown in Fig. \ref{max_K_figure}, the maximum K number decreases monotonically with increasing distance $R$ between the transmitting and receiving arrays, and the rate of decrease diminishes monotonically as well.
The maximum K number also decreases monotonically with increasing $\theta$. When $\theta = 0$, i.e., when the transmitting and receiving arrays are directly facing each other, the maximum K number reaches its peak.
Additionally, it is observed that the maximum K number calculated via (\ref{K2_max}) closely matches the maximum K number obtained numerically, particularly when $R > 300\lambda$. This validates the accuracy of the approximate expression for the K number provided by (\ref{K2}) and (\ref{K2_max}).

\subsection{K Number and EDoF}
Next, we explore the relationship between the K number and the EDoF of the MIMO channel. Let the antenna spacing of the transmitting array $\mathcal{L}_s$ and the receiving array $\mathcal{L}_p$ be $\Delta_s$ and $\Delta_p$, respectively. Consequently, the number of transmitting antennas and receiving antennas are $N_t = L_s / \Delta_s + 1$ and $N_r = L_p / \Delta_p + 1$, respectively. Let the MIMO channel be represented by $\mathbf{H} \in \mathbb{C}^{N_r \times N_t}$. Considering the near-field spherical wavefront effect \cite{tang2023line}, \cite{do2022line}, the complex gain between the $n_r$-th receiving antenna and the $n_t$-th transmitting antenna can be modeled as
\begin{equation}
h_{n_r,n_t}=\left[ \mathbf{H} \right] _{n_r,n_t}=\frac{\lambda}{4\pi r_{n_r,n_t}}e^{\mathrm{j}k_0r_{n_r,n_t}},
\end{equation}
where $r_{n_r, n_t}$ denotes the distance between the $n_r$-th receiving antenna and the $n_t$-th transmitting antenna, with $n_r = 1, \ldots, N_r$ and $n_t = 1, \ldots, N_t$.

\begin{figure}[htbp]
    \centering
    \subfigure[$\Delta_s = \Delta_p = \lambda /2$] {\label{KSV_A}
        \includegraphics[scale=0.3]{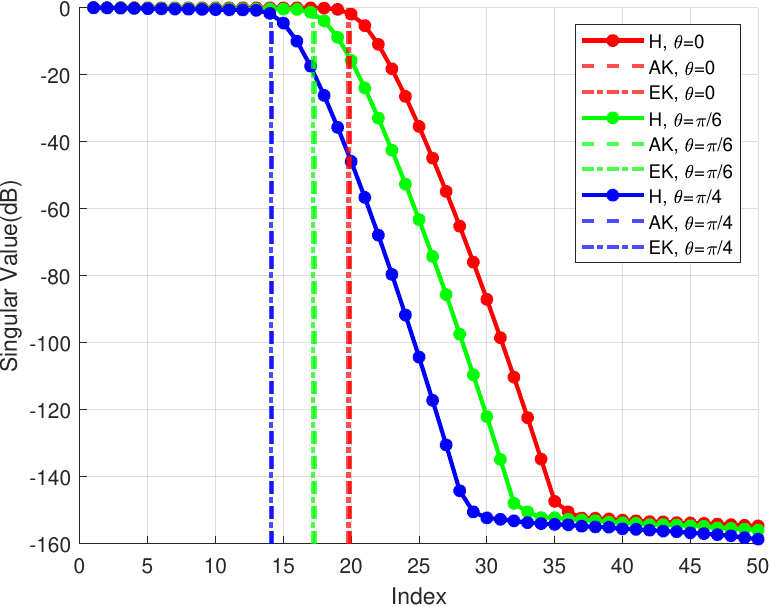}
    }
    \subfigure[$\Delta_s = \Delta_p = \lambda /4$] { \label{KSV_B}
        \includegraphics[scale=0.3]{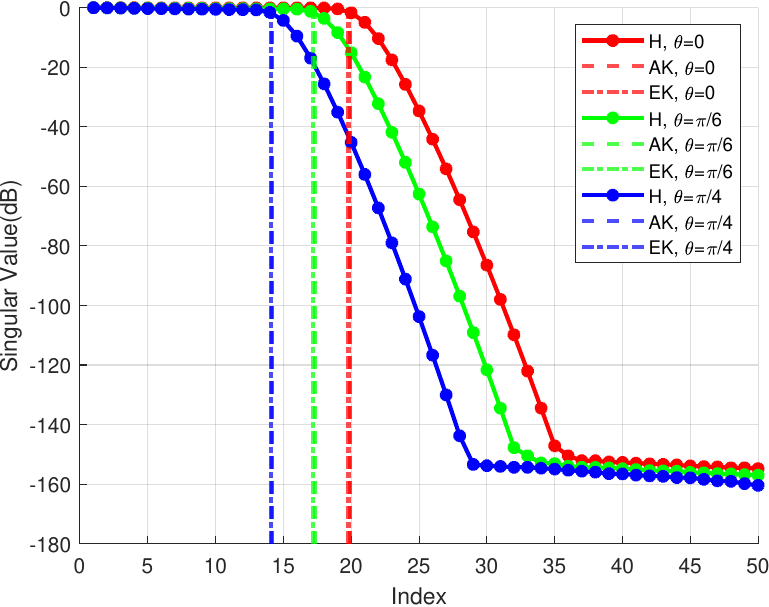}
    }
    \caption{The relationship between the singular values of the LoS MIMO channel and the K number for different positions of the receiving array.
    \label{KSV}
}
\end{figure}

Figure \ref{KSV} illustrates the distribution of normalized singular values of the channel $\mathbf{H}$ when $L_s = L_p = 100 \lambda$, $R = 500 \lambda$, and $\lambda = 0.01$ m. Figs. \ref{KSV_A} and \ref{KSV_B} correspond to $\Delta_s = \Delta_p = \lambda /2$ and $\Delta_s = \Delta_p = \lambda /4$, respectively. The meanings of ``AK" and ``EK" are the same as in Fig. \ref{max_K_figure}. ``H" represents the distribution of singular values of the MIMO channel matrix under the receiving direction corresponding to ``EK". Let the ordered singular values of $\mathbf{H}$ be denoted as $\sigma_1 \ge \sigma_2 \ge \cdots \ge \sigma_N$. As seen in Fig. \ref{KSV_A}, regardless of the position of the receiving array $\mathcal{L}_p$, the rate of decrease of $\sigma_n$ can always be divided into two stages: in the first stage, the singular values $\sigma_n$ remain roughly constant until reaching a critical threshold, after which $\sigma_n$ rapidly decays. In fact, this ``threshold" is referred to as the EDoF, which is related to the spatial multiplexing capability of MIMO communication \cite{liu2024near}.

Fig. \ref{KSV_A} also shows that ``EK" can effectively delineate the two stages of singular value decay, validating that the K number is a good approximation of the EDoF. Additionally, we observe that ``AK" and ``EK" almost overlap, which validates the correctness of (\ref{K2}) and (\ref{K2_max}). Meanwhile, we observe that for the same $R$, the larger the $\theta$, the smaller the K number and the EDoF of $\mathbf{H}$. This conclusion aligns with the properties of local spatial bandwidth shown in Fig. \ref{max_BW_figure} and Fig. \ref{max_K_figure}.

Comparing Fig. \ref{KSV_A} and Fig. \ref{KSV_B}, it can also be seen that as long as the dimensions of the transmit and receive arrays remain unchanged, adjusting the antenna spacing (or the number of antennas) does not affect the K number or the EDoF of the LoS channel. Therefore, the key factor influencing the EDoF of the array is not the number of antennas, but the size of the array.

\begin{figure}[htbp]
    \centering
    \subfigure[$\varphi'=\pi/2$] {
    \label{KSV_C}
        \includegraphics[scale=0.3]{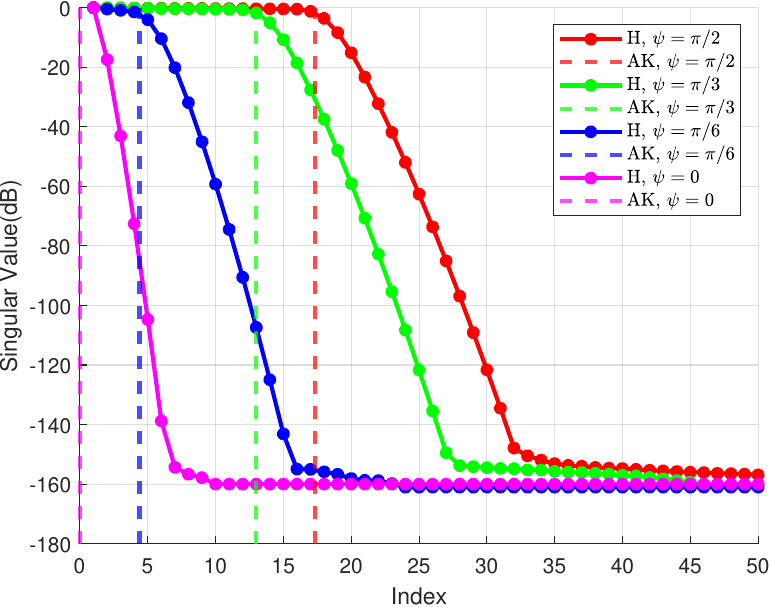}
    }
    \subfigure[$\psi=\pi/2$] { 
    \label{KSV_D}
        \includegraphics[scale=0.3]{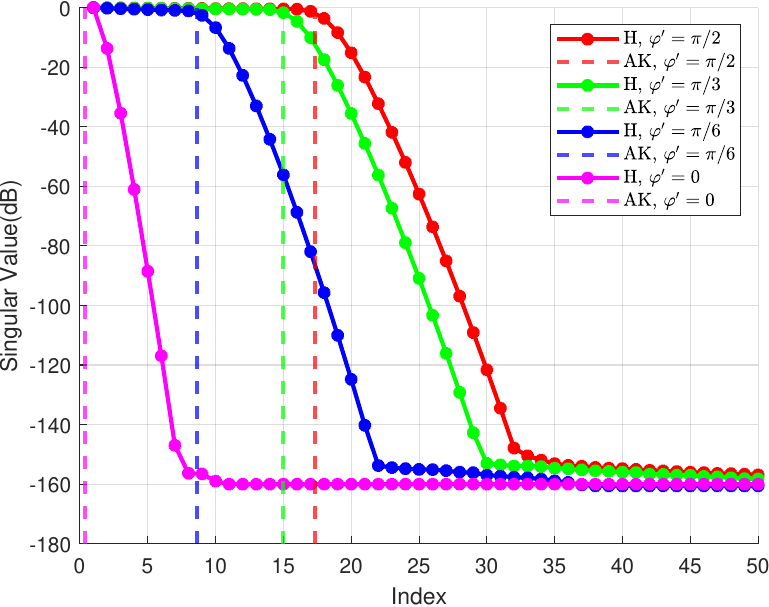}
    }
    
    \caption{The relationship between the singular values of the LoS MIMO channel and the K number for different orientations of the receiving array.
    \label{KSV_2}
}
\end{figure}

Fig. \ref{KSV_2} illustrates the relationship between the normalized singular value distribution of the channel matrix $\mathbf{H}$ and the orientation of the receiving array $(\psi, \varphi')$. The parameters are set as $L_s = L_p = 100 \lambda$, $(R, \theta) = (500\lambda, \pi/6)$, $\Delta_s = \Delta_p = \lambda/2$, and $\lambda = 0.01 \mathrm{m}$.

Fig. \ref{KSV_C} examines the EDoF of the channel matrix $\mathbf{H}$ when $\varphi' = \pi/2$ and $\psi$ varies in the range $[0, \pi/2]$. ``H" denotes the singular value distribution of the channel matrix $\mathbf{H}$ for the receiving array orientation $(\psi, \varphi')$. ``AK" represents the approximate K number calculated using (\ref{K2}). It can be observed that when $\psi = \pi/2$, both the K number and EDoF of the channel reach their maximum values, indicating that $\mathcal{L}_s$ and $\mathcal{L}_p$ are coplanar. Conversely, when $\psi = 0$, the K number and EDoF are minimized, corresponding to the situation where $\mathcal{L}_p$ is perpendicular to $\mathcal{L}_s$. 

Similarly, Fig. \ref{KSV_D} investigates the EDoF of the channel matrix $\mathbf{H}$ when $\psi = \pi/2$ and $\varphi'$ varies in the range $[0, \pi/2]$. It is found that the K number and EDoF of the channel reach their maximum values when $\varphi' = \pi/2$, indicating that $\mathcal{L}_p$ is perpendicular to the centerline of the transmit and receive arrays. When $\varphi' = 0$, the K number and EDoF are minimized, indicating that $\mathcal{L}_p$ is collinear with the centerline of the transmit and receive arrays.

% \begin{figure}[htbp]
%     \centering
%     \subfigure[$\varphi'=\pi/2$] {
%     \label{KSV_C}
%         \includegraphics[scale=0.3]{KSV_v_psi.pdf}
%     }
%     \subfigure[$\psi=\pi/2$] { 
%     \label{KSV_D}
%         \includegraphics[scale=0.3]{KSV_v_phi.pdf}
%     }
    
%     \caption{The relationship between the singular values of the LoS MIMO channel and the K number for different orientations of the receiving array.
%     \label{KSV_2}
% }
% \end{figure}

% \begin{figure*}[!t]
% \centering
% %\includegraphics[width=3in]{fig5}
% \subfloat[subfig figure title]{
% 		\includegraphics[scale=0.4]{KSV_L500_025L.pdf}}
% \subfloat[subfig figure title]{
% 		\includegraphics[scale=0.4]{KSV_L500_05L.pdf}}
% \caption{title}
% \label{KSV}
% \end{figure*}

\section{Conclusion}
In this paper, we derive a closed-form expression for the spatial bandwidth of LoS channels with linear LSAA in 3D space. This expression provides rich geometric insights into the spatial bandwidth and the EDoF of LoS channels. It is observed that when the transmit and receive arrays are coplanar and the receive array is perpendicular to the line connecting the centers of the transmit and receive arrays, the spatial bandwidth at the center of the receive array is maximized. In this case, the LoS channel achieves nearly the maximum EDoF for spatial multiplexing.

\section*{Acknowledgement}
This work was supported in part by the National Natural Science Foundation of China under Grant No. 62371123, Chongqing Natural Science Joint Fund Project under Grant No. CSTB2023NSCQ-LZX0121, ZTE Industry-University-Institute Cooperation Funds under Grant No. IA20231212016, Jiangsu Province Science and Technology Project under Grant BE2021031, and Research Fund of National Mobile Communications Research Laboratory Southeast University under Grant No.2023A03.

\appendices
% \section{Proof of the Proposition \ref{prop1}}
% According to (\ref{local_BW_def}) and letting $I=\left[ -\frac{\alpha}{2}+\beta ,\frac{\alpha}{2}+\beta \right] $, we have
% \begin{equation}
% \label{proof1}
% \begin{aligned}
% & \omega(\psi, \varphi+\pi ; \alpha, \beta) \\
% & =\max _{\gamma \in I} f(\psi, \varphi+\pi, \gamma)-\min _{\gamma \in I} f(\psi, \varphi+\pi, \gamma) \\
% & =\max _{\gamma \in I}-f(\psi, \varphi, \gamma)-\min _{\gamma \in I}-f(\psi, \varphi, \gamma) \\
% & =-\min _{\gamma \in I}f(\psi, \varphi, \gamma)+\max_{\gamma \in I}f(\psi, \varphi, \gamma) \\
% & =\omega(\psi, \varphi ; \alpha, \beta).
% \end{aligned}
% \end{equation}

\section{Proof of the Proposition \ref{prop2}}
Since (\ref{w_closed_form_expr}) is a piecewise function, $\omega^{\max}$ is determined by the maximum value of each segment. According to (\ref{periodic_function}),
\begin{equation}
    \omega ^{\max}=\max_{\psi \in \left[ 0,\pi \right] , \varphi' \in \left[ 0,\pi \right]} \,\,\omega \left( \psi ,\varphi';\alpha ,\beta \right).
\end{equation}

When $0 \le \varphi' \le \alpha /2$, we have
\begin{equation}
\label{seg1}
\omega \left( \psi ,\varphi' \right) \le k_0\left( 1-\cos \alpha \right), 
\end{equation}
the maximum value of (\ref{seg1}) is attained when $(\psi,\varphi') = (\pi /2,\alpha/2)$.

When $\alpha/2 \le \varphi' \le \pi - \alpha /2$, we have
\begin{equation}
\label{seg2}
\omega \left( \psi ,\varphi' \right) \le 2k_0\sin \left( \alpha /2 \right) , 
\end{equation}
the maximum value of (\ref{seg2}) is attained when $(\psi,\varphi') = (\pi /2,\pi /2)$.

When $\pi - \alpha /2 \le \varphi' \le \pi$, we have
\begin{equation}
\label{seg3}
\omega \left( \psi ,\varphi' \right) \le k_0\left( 1-\cos \alpha \right), 
\end{equation}
the maximum value of (\ref{seg3}) is attained when $(\psi,\varphi') = (\pi /2,\pi- \alpha/2)$. Thus,
\begin{equation}
\omega ^{\max}=k_0\cdot \max \left\{ 1-\cos \alpha ,2\sin \left( \alpha /2 \right) \right\} =2k_0\sin \left( \alpha /2 \right). 
\end{equation}
Note that at this point, $\hat{\boldsymbol{v}} = \hat{\boldsymbol{v}}_{NP}$.

\bibliographystyle{IEEEtran}
\bibliography{Mybib}
\end{document}